%% file: brane.tex
\documentstyle[12pt,psfig]{article}

\newcommand\Tr{{\rm Tr}}
\def\epsfig{\psfig}
\begin{document}
\thispagestyle{empty}
\rightline{LMU-HEP-97-16}
\rightline{MRI-PHY/970614}
\rightline{hep-th/9706057}
\vspace{2 truecm}

\centerline{\bf 
AN ORIENTIFOLD OF THE SOLITONIC FIVEBRANE}

\vspace{1.2truecm}
\centerline{\bf Stefan F\" orste, Debashis Ghoshal and
Sudhakar Panda\footnote{\noindent Permanent Address:
Mehta Research Institute of Mathematics \&\ Mathematical
Physics, Chhatnag Road, Jhusi, Allahabad 211506, India}\footnote{
\noindent
E-mail: \tt Stefan.Foerste@physik.uni-muenchen.de\hfill\\ 
\hphantom{E-mail: }Debashis.Ghoshal@physik.uni-muenchen.de\hfill \\ 
\hphantom{E-mail: }panda@mri.ernet.in}}
\vspace{.5truecm}
\centerline{\em Sektion Physik, Universit\"at M\"unchen}
\centerline{\em Theresienstra\ss e 37, 80333 M\"unchen, Germany}

\vspace{2.2truecm}


\vspace{.5truecm}

\begin{abstract}
We study an orientifold of the solitonic fivebrane of type II
string theory. The consideration is restricted to a space-time
domain which can be described by an exact conformal field 
theory. There are no IR divergent contributions to tadpole
diagrams and thus no consistency conditions arise.
However, extrapolating the results to spatial infinity leads to
consistency conditions implying that there are four (physical)
D-6-branes sitting at each of the two orientifold fixed planes.   
\end{abstract}
\vfill

\newpage

\section{Introduction}

One of the first solitonic solutions to be found in superstring theory is a
fivebrane that carries the magnetic charge of the rank two 
antisymmetric 
gauge potential\cite{chs}. This is dual to the elementary string itself 
which is electrically charged. The fivebrane configuration is a 
solution of the equations of motion derived from the low energy effective
action. Although it is constructed as a classical solution, 
the fivebrane breaks half of the
spacetime supersymmetries and hence its energy and charge saturate 
the BPS bound. This makes the fivebrane stable. Since the gauge
potential arises in the Neveu-Schwarz sector, this solution is called
an NS fivebrane. 

In the last couple of years there has been a fair amount of progress in our 
understanding of the non-perturbative aspect of superstrings (see
Ref.\cite{review} for some recent reviews). One finds extended 
solitonic objects, $p$-branes, of all dimensions that couple to the variety of 
Ramond-Ramond (RR) gauge potentials in type II strings\cite{polch}.
These $p$-branes are topological defects arising from open strings 
that have Dirichlet boundary condition in $(9-p)$ directions and hence 
are tied to a $p$ dimensional subspace at all times\cite{horava}\cite{dlp}.
The end-points of the open strings thus sweep the worldvolume of
the D(irichlet)-$p$-branes. Various properties of D-branes can 
therefore be computed using perturbation theory of open strings
with appropriate modification of boundary conditions\cite{dreview}. 

Another kind of topological defect arises when one generalizes 
space-time orbifold by combining it with the world-sheet parity of 
type II strings\cite{horvifold}\cite{dlp}. The {\it orientifold}  hyperplane
obtained this way is not a dynamical object (at least in perturbation
theory of open strings) but nevertheless carries charge of an 
appropriate RR field. Near the orientifold, unoriented string diagrams 
contribute. Indeed type I string itself can be obtained by quotienting
the ten dimensional type IIB theory by the world-sheet
parity. More general examples of this type provide new string vacua
in lower dimensions\cite{gp}\cite{otherorientifolds}.

In this paper we would like to study an orientifold of type II string in
the background of an NS fivebrane. The fivebrane
modifies the flat geomtery of the space transverse to it. The
orientifold construction therefore needs to be carried out in a
curved geometry. What makes this tractable is that there exists
an exact description of the NS fivebrane as a superconformal field
theory in some region of spacetime. We analyze the spectrum and 
compute the one-loop tadpole and look for possible divergence
as a signal of inconsistency. Such divergences appear due to
massless fields. In a consistent theory the total divergence from
all diagrams must cancel to have conservation of charge. This requires
to have D-branes with specific properties. Surprisingly
we will not encounter any real divergence. This would imply that  an 
orientifold of the NS fivebrane is always consistent. However subtleties
arise due to the curved background and nonconstant dilaton.  We will 
discuss this issue in more detail later. 

One motivation to look at this problem came from the `derivation'
of the Seiberg-duality \cite{sireview}\ in $N=1$ supersymmetric 
gauge theories using simple manipulations with branes\cite{egk}. 
The duality for orthogonal and symplectic gauge groups require one to 
have NS fivebranes and orientifolds simultaneously\cite{ejs}\cite{egkrs}. In 
any case, the problem is interesting in its own right as one applies the 
orientifold construction to a non-trivial conformal field theory. As far as we 
know the orientifold technique has been applied in detail to flat spacetime 
only.  

\section{Geometry and Conformal Field Theory of the NS Fivebrane}

\begin{figure}
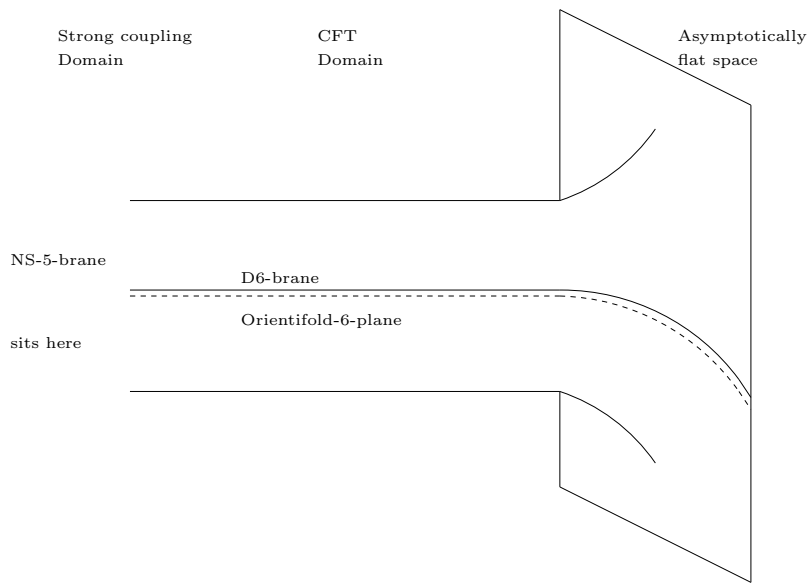
  
\begin{center}
\input fig.pstex_t
\end{center}
\caption{The geometry of the NS-5-brane orientifold.}
\label{worm}
\end{figure}

The worldvolume of the fivebrane spans six space-time dimensions and
is transverse to the remaining four. Let the fivebrane lie along the 
coordinates $\{x^0,x^1,\cdots,x^5\}$, and $\{x^6,x^7,x^8,x^9\}$ be the 
transverse space. Let us recall the explicit solution for the metric, 
dilaton and the antisymmetric tensor field \cite{chs}:
\begin{eqnarray}
ds^2 &= & {\displaystyle\sum_{\mu,\nu=0}^5} \eta_{\mu\nu}dx^\mu dx^\nu +
e^{-2\varphi}\left(dr^2 + r^2d\Omega_3^2\right)\nonumber\\
e^{-2\varphi} &= & e^{-2\varphi_0}\left(1 + {k\over r^2}\right)\label{five}\\
H &= & -\, 2k d\Omega_3\nonumber
\end{eqnarray}
The geometry of the transverse space is modified by the presence 
of the brane. It is still asympotically flat, but as we approach the 
`position' of the brane, distances stretch out resembling a 
semi-wormhole throat. This part of space is cylindrical with a 
three-sphere $S^3$ as its base (see figure \ref{worm}). 
The string coupling gets stronger along the length of 
the cylinder as the dilaton field is linear in this direction, 
and finally diverges at the core of the brane. 

In the region of the wormhole throat, the fivebrane is exactly 
described by a superconformal field theory. The SCFT relevant 
for the transverse coordinates is an $N=4$ supersymmetric $SU(2)$
WZW model tensored with a free scalar field with a background 
charge (a Feigin-Fuchs model). The level of the WZW model is 
the charge of the NS fivebrane (upto a factor of $\alpha'$), which
is also the size of the $S^3$. The total central charge is exactly six
(due to supersymmetry), and this determines the background
charge of the Feigin-Fuchs field. The fermions are essentially
free. For more details, see Ref.\cite{chs}. 

The SCFT in
the longitudinal directions is a theory of free bosons and fermions
with $c=9$. The bosonic part of the $SU(2)_k$ WZW action is, (see
Ref.\cite{difra} for an excellent text on CFT in general and WZW 
models in particular),
\begin{equation}\label{wzws}
S_{WZW} = {k\over 8\pi}\int_{\partial{\cal B}} 
\Tr(\partial g^{-1}\bar\partial g)
- {ik\over 12\pi}\int_{\cal B} \Tr(g^{-1}dg)^3 
\end{equation}
where ${\cal B}$ is a three manifold with the string worldsheet 
as its boundary. As is clear from the solution (\ref{five}) the Wess-Zumino
term in action (\ref{wzws}) describes the coupling of the fivebrane
to the NS antisymmetric tensor. 

There are two conserved currents
\begin{equation}\label{currents}
J(z)=(\partial g)g^{-1}\hbox{\rm \ and\ }\bar{J}(\bar{z})=g^{-1}\bar\partial g
\end{equation}
corresponding to the Kac-Moody gauge symmetries
\begin{equation}\label{kac}
g(z,\bar z)\rightarrow \Lambda(z)g(z,\bar z){\bar\Lambda}^{-1}(\bar z)
\end{equation}
The central charge is given in terms of the level $c_{WZW}=3k/(k+2)$. 
The Feigin-Fuchs part is a free boson with a background charge
$Q$ induced by the linear dilaton of the fivebrane. The central
charge shifts to $c_{FF} = 1+3Q^2$ with $Q=\sqrt{2/(k+2)}$
(the $\alpha'\sim1/k$ corrections have been taken into account here). The
total bosonic contributions from the WZW and FF part add up to
four, equal to that of four flat coordinates.

This model has a discrete symmetry, whose action is  
the worldsheet parity transformation $\Omega$ combined with a 
`reflection' $R$ taking the group element to its inverse:
\begin{equation}\label{symmetry}
\Omega R : \begin{array}{l}
g(z,\bar z)\; \rightarrow\; g^{-1}(\bar z,z)\\
X(z,\bar z)\rightarrow X(\bar z,z)
\end{array}
\end{equation}
where $X(z,\bar z)$ stands for the other bosons. 
The first term in the WZW action (\ref{wzws}) is invariant separately
under the action of $\Omega$ and $R$ whereas the Wess-Zumino term
is invariant only under the combined action. So (part of) the NS 
antisymmetric tensor survives the $\Omega R$ 
projection\footnote{This is unlike the type I string obtained by an $\Omega$ 
projection of type IIB theory where the NS antisymmetric tensor gets
eliminated.}.
Notice that under this symmetry the Kac-Moody currents are interchanged
\begin{equation}\label{moody}
J(z) \leftrightarrow -\, {\bar J}(\bar z)
\end{equation}
which is consistent with Eq.(\ref{kac}). The orientifold of the fivebrane
we want to study is obtained by gauging the symmetry 
$\Omega R$ in Eq.(\ref{symmetry}).

The transformation (\ref{symmetry}) leaves two fixed points on the $SU(2)$
group manifold $S^3$. Hence the resulting orientifold plane is extended
along all directions parallel to the fivebrane and one direction transverse
to it, namely the Feigin-Fuchs coordinate.  We therefore obtain an 
orientifold 6-plane. This carries charge of the RR 7-form gauge potential, 
which is part of the spectrum of type IIA string. This charge can be 
neutralized by adding D-6-branes parallel to the orientifold plane. 
Taking some of the directions ($x^1,\cdots,x^5$) to be compact and
performing T-duality on them, one can go back and forth from IIA to IIB 
theory. 

As can be seen from (\ref{symmetry}) the two fixed points are in the 
centre of $SU(2)$ $g=\pm{\bf 1}$. There is another symmetry in the WZW
theory\cite{black}\cite{trivial}, namely 
$\Omega R':\, g(z,\bar z)\rightarrow -g^{-1}(\bar z,z)$.
This results in a fixed $S^2$ surface at the equator of $S^3$.
This would lead to orientifold 8-surfaces and D-8-branes wound on the 
$S^2$, but we will not discuss this case any farther.

\section{Massless Spectrum of the Fivebrane Orientifold}

Before going into the details of the spectrum, let us point out one
subtlety. As described, for example in \cite{diku}, the presence of 
the linear dilaton shifts the mass square by a term proportional 
to the square of the background charge. This can be understood 
by dividing the vertex operator by the string coupling `constant'.
Down the throat of the wormhole the string coupling increases, and
particles which are massless at asymptotic regions acquire an
effective mass. The notion of the massless spectrum to be
discussed presently is understood with this caveat. 

The fivebrane breaks the ten dimensional little group $SO(8)$ to
the six dimensional little group $SO(4)\sim SU(2)\times SU(2)$ 
times $SU(2)$ of the WZW part of the transverse space. We will
label the states by their six dimensional Lorentz quantum numbers.  
The massless 
spectrum of type IIA theory is constructed out of an ${\bf 8}_v$ from 
the left and right NS-sector, an ${\bf 8}_s$ from the left R-sector,
and an ${\bf 8}_c$ from the right R-sector. These $SO(8)$ representations 
decompose as follows
\begin{eqnarray}
{\bf 8}_v & \rightarrow & ({\bf 2},{\bf 2}) + 4({\bf 1},{\bf 1}) \nonumber \\
{\bf 8}_s & \rightarrow & 2({\bf 2},{\bf 1}) + 2({\bf 1},{\bf 2}) \label{decomp} \\
{\bf 8}_c & \rightarrow & 2({\bf 2},{\bf 1}) + 2({\bf 1},{\bf 2}) . \nonumber
\end{eqnarray}
Taking $\Omega R$ invariant combinations from the left and right
sectors, we arrive at the bosonic spectrum
\begin{eqnarray}
\mbox{NS-NS}: &\mbox{\ } &  ({\bf 3},{\bf 3}) + 4({\bf 2},{\bf 2}) 
+ 11 ({\bf 1},{\bf 1}) \\
\mbox{R-R} : &\mbox{\ } & ({\bf 3},{\bf 1}) + ({\bf 1},{\bf 3}) + 4({\bf 2},{\bf 2}) 
+ 6({\bf 1},{\bf 1}) .
\end{eqnarray}
A part of the NS antisymmetric tensor field survives as vectors and scalars. 
Together with the fermions these states combine into the gravity multiplet
and four vector multiplets of $N=4$ non-chiral supergravity in
six dimensions.

As stated in Ref.\cite{gp} $\Omega R$ does not give rise to twisted sector 
states but may require the inclusion of open strings in order to obtain
a consistent theory. These open strings can have NN, DN, ND or DD
boundary conditions. The oscillator parts in the  mode expansion of a
free boson and fermion are\cite{dreview}
\begin{eqnarray}
\partial X(\sigma, 0) & = & {\displaystyle\sum_m} \alpha_m
\left(e^{im\sigma} \pm e^{-im\sigma}\right)\\
\psi(\sigma,0) & = & {\displaystyle\sum_r} \psi_r e^{ir\sigma}\\
\tilde\psi(\sigma,0) & = & \pm {\displaystyle\sum_r} \psi_r e^{-ir\sigma}
\end{eqnarray}
where the $\pm$ refer to NN or DD boundary conditions for  integer
moded $\alpha$, and to ND or DN for half-integrally moded ones. 
Under worldsheet parity $\Omega:\sigma\rightarrow\pi - \sigma$,
the oscillators transform as
\begin{equation}\label{alpha}
\alpha_m\rightarrow \pm e^{i\pi m}\alpha_m\;\;\;\mbox{and}\;\;\;
\psi_r\rightarrow \pm e^{i\pi r} \psi_r.
\end{equation}
In the WZW sector the role of the $U(1)$ current is played by
the Kac-Moody currents Eq.(\ref{kac}). The $R$ action gives an additional
sign in the bosons and fermions of the WZW sector. Here we should 
point out that our treatment for the WZW sector of open strings is
purely algebraic. While it would be nice to have further insight the
algebraic viewpoint suffices for our present purpose. 

In the open string spectrum, one gets a vector
\begin{equation}
\psi^\mu_{-{1\over 2}} |0,\alpha\beta\rangle \lambda_{\beta\alpha},\;\;\;\;\;
\lambda = \mp \gamma_{\Omega R}\lambda^T\gamma^{-1}_{\Omega R}
\label{vector}
\end{equation}
where the $\mp$ refers to DD or NN boundary conditions respectively. 
Additionally one gets scalars from the WZW and FF fermions
\begin{eqnarray}
\psi^{WZW}_{-{1\over 2}} |0,\alpha\beta\rangle \lambda_{\beta\alpha},&\;\;\; &
\lambda = \pm \gamma_{\Omega R}\lambda^T\gamma^{-1}_{\Omega R}\\
\psi^{FF}_{-{1\over 2}} |0,\alpha\beta\rangle \lambda_{\beta\alpha},&\;\;\; &
\lambda = \mp \gamma_{\Omega R}\lambda^T\gamma^{-1}_{\Omega R}.
\end{eqnarray}
The restriction on the Chan-Paton matrices arise for open strings with
end points stuck at the orientifold fixed plane. Otherwise $\Omega R$
maps them to their images and puts no further condition. 
To be able to form a supermultiplet one needs to have the same projections
for the four scalars and the vector. For consistency therefore we choose
DD conditions in the WZW sector and NN for the rest.  This results in 
D-6-branes parallel to the orientifold. 

\section{One Loop Calculation}
 
Now we come to the computation of the one loop amplitudes. The general
framework is discussed at length in Ref.\cite{gp} which we will follow 
closely. We will only point out special features of our model and the
reader is referred to \cite {gp} for additional details. Let us begin by
recalling the expression for the one loop amplitude
\begin{equation}
\int {dt\over 2t}\left(\Tr_C \left( (-)^{\scriptstyle{\bf F}}\, 
\mbox{\bf P} e^{-2\pi t(L_0+{\tilde L}_0)}\right) 
+ \Tr_O\left( (-)^{\scriptstyle{\bf F}}\, \mbox{\bf P} e^{-2\pi t L_0}\right)\right)
\label{loop}
\end{equation}
where $\mbox{\bf P}={1+\Omega R\over 2}\, {1 + (-)^F\over 2}$ is the
orientifold and GSO projections and $\mbox{\bf F}$ is the spacetime
fermion number. The traces $\Tr_O$ and $\Tr_C$ refer to worldsheets
with or without boundaries respectively. Diagrams with an $\Omega$ 
insertion correspond to nonorientable surfaces, the M\" obius strip and
the Klein bottle; and those without are the cylinder and the torus
respectively. Since modular invariance of the torus removes the region
of $t$ integration that would lead to IR divergence, we need to 
consider the other three diagrams only. 

The ingredients needed for the computation are the characters of the
free bosonic and fermionic fields and the bosonic WZW model. The 
nontrivial part is the character of the $SU(2)_k$ WZW model \cite{difra}
\begin{equation}
\chi_j(q) = {q^{(2j+1)^2/4(k+2)}\over \eta^3(q)}
\displaystyle\sum_{n=-\infty}^\infty\left( 2j+1+2n(k+2)\right)
q^{n\left(2j+1+(k+2)n\right)}\label{charac}
\end{equation}
where $(2j+1)$ is the dimension of the spin $j$ representation of $SU(2)$.
In the WZW model $j$ takes values from $0$ to $k/2$. When taking
the trace in Eq.(\ref{loop}), one has to sum over the representations. This
is the analog of the sum over momenta and windings in the compact 
coordinates in Ref.\cite{gp}. 

The $\Omega R$ projection in the Klein bottle implies that we only keep
excitations that are identical in the left and the right sectors. On the $SU(2)$
quantum numbers this puts the restriction $\tilde j = j, \tilde m = -m$ 
relating the ground states of the left and the right sector\footnote{For the 
other symmetry $\Omega R'$, the analogous condition is 
$\tilde j = j, \tilde m = m$.}. This results in the following expression
\begin{equation}
(1-1){V_6 L\over 8}\int_0^\infty {dt\over t^4\sqrt{t}}\; 
{f^8_4(e^{-2\pi t})\over f^5_1(e^{-2\pi t})}
\displaystyle\sum_{j=0}^{k/2} \chi_j(e^{-4\pi t})\label{klein}
\end{equation}
where $V_6$ and $L$ are the regularized volume of the noncompact space
and the Feigin-Fuchs direction. The functions $f_i(q)$ are
\begin{eqnarray}
f_1(q) = q^{1/12}\displaystyle\prod_{n=1}^\infty (1 - q^{2n}),& \;\;\; &
f_2(q) = \sqrt{2}q^{1/12}\displaystyle\prod_{n=1}^\infty (1 + q^{2n}),\\
f_3(q) = q^{-1/24}\displaystyle\prod_{n=1}^\infty (1 + q^{2n-1}),& \;\;\; &
f_4(q) = q^{-1/24}\displaystyle\prod_{n=1}^\infty (1 - q^{2n-1}).
\end{eqnarray}
(The function $\eta(q)$ in (\ref{charac})  is the same as $f_1(\sqrt{q})$.) These
functions have the modular property
\begin{equation}\label{modular}
f_1(e^{-\pi/t}) = \sqrt{t} f_1(e^{-\pi t}),\;\;\;
f_2(e^{-\pi/t}) = f_4(e^{-\pi t}),\;\;\;
f_3(e^{-\pi/t}) = f_3(e^{-\pi t}).
\end{equation}
The corresponding expression for the M\" obius strip is
\begin{equation}
-\; (1-1) {V_6 L\over 64\sqrt{2}}
\sum\limits_{I=1,2}\Tr(\gamma^{-1}_{I,\Omega R}\gamma^T_{I,\Omega R})
\int_0^\infty {dt\over t^4\sqrt{t}}\; 
{f^8_2(e^{-\pi t + {i\pi\over 2}})\over f^5_1(e^{-\pi t + {i\pi\over 2}})}
\displaystyle\sum_{j=0}^{k/2} 
e^{-{i\pi(2j+1)^2\over 4(k+2)}}\chi_j(e^{-2\pi t + i\pi})\label{mobius}
\end{equation}
(Notice that in the open string sector the zero modes $J_0$ have 
$\Omega R$ eigenvalue +1 due to DD boundary conditions.)
The shift in the argument of the characters comes from the fact that 
the $\Omega R$ insertion is equivalent to an $e^{i\pi L_0}$ insertion.
Finally the cylinder diagram contributes
\begin{equation}
(1-1) {V_6 L\over 64\sqrt{2}} 
\sum\limits_{I=1,2}\left(\Tr \gamma_{I,\scriptstyle{\bf 1}}\right)^2
\int_0^\infty {dt\over t^4\sqrt{t}}\; 
{f^8_4(e^{-\pi t })\over f^5_1(e^{-\pi t })}
\displaystyle\sum_{j=0}^{k/2} \chi_j(e^{-2\pi t })\label{cylinder}
\end{equation}
We only consider D-6-branes that are stuck to the orientifold planes. 

To extract the IR behaviour (the $t\to 0$ limit), the following identities 
pertaining to the characters of the WZW model are useful
\begin{eqnarray}
e^{-{\pi (2j+1)^2 t\over (k+2)}}&\!\!\displaystyle\sum_{n=-\infty}^\infty& 
\!\!\left( 2j+1+2(k+2)n\right)\, e^{-4\pi tn\left(2j+1+(k+2)n\right)}\nonumber\\
=\; {1\over\sqrt{4t^3(k+2)}} &\!\!\displaystyle\sum_{n=1}^\infty&
\!\! n e^{-{\pi n^2\over 4t(k+2)}} 
\sin\left({\pi (2j+1)n\over (k+2)}\right)\label{one}\\
e^{-{\pi (2j+1)^2 t\over 2(k+2)}}&\!\!\displaystyle\sum_{n=-\infty}^\infty& 
\!\!\left( 2j+1+2(k+2)n\right)\, e^{-2\pi (t+{i\over 2})n\left(2j+1+(k+2)n\right)} 
\nonumber\\
=\; {1\over\sqrt{8t^3(k+2)}} &\!\!\displaystyle\sum_{n=1}^\infty&
\!\! n e^{-{\pi n^2\over 8t(k+2)}} \sin\left({\pi (2j+1)n\over 2(k+2)}\right)
\left(1 - e^{i\pi(k+2j+n)}\right)\label{two}
\end{eqnarray}
The above formulas are easily obtained by Poisson resummation. 

The length of the Klein bottle, M\" obius strip and cylinder are related
to the loop parameter $t$ by $t={1\over 4l},\, {1\over 8l},\, {1\over 2l}$
respectively\cite{gp}. Taking this into account and  using
Eqs.(\ref{modular}), (\ref{one}) and (\ref{two}), we obtain the leading 
IR contribution
\begin{equation}\label{convergence}
\int\limits_{l\to\infty}dl\, \left(64 - 
8\sum\limits_{I=1,2}\Tr(\gamma^{-1}_{I,\Omega R}\gamma^T_{I,\Omega R})
+{1\over 2} \sum\limits_{I=1,2}\left(\Tr \gamma_{I,\scriptstyle{\bf 1}}\right)^2\right)
e^{-\pi l/3}.
\end{equation}
In the above we have taken the value $k=1$ corresponding to the
singly charged NS fivebrane. The expression (\ref{convergence}) is
clearly convergent as $l\to \infty$, but the argument of the exponential is
proportional to the shift in mass due to the linear dilaton. In the 
asymptotic region the dilaton approaches a constant value and there are
massless particles in the spectrum. The one loop amplitude is then
truly divergent. We assume that the results of the calculation 
carried out in the region of the wormhole throat can be extrapolated
to the asymptotically flat region. 

If the branes are distributed equally at the two fixed points, the
`divergence' can be cancelled for four physical D-6-branes with 
Chan-Paton
factors transforming in $SO(8)$ at each orientifold plane. If D-6-branes 
away
from the fixed planes are allowed, then they contribute additional 
divergence that cannot be cancelled. This fits nicely with the fact that
the $\Omega R$ projection does not put any constraint on such branes
since it maps them to their images.  The final result is that we get four
(physical) D-6-branes stuck at each of the two orientifold 6-planes leading
to an $SO(8)\times SO(8)$ gauge symmetry. 

\section{Summary}

In this paper we have studied an orientifold of the conformal field theory
of the NS fivebrane. Unlike the orientifolds of flat spacetime, we do not 
encounter any real IR divergence in the one loop tadpole diagrams. 
However in the curved background of the fivebrane the asymptotically
massless particles acquire an effective mass induced by the linear
dilaton field. Extrapolating the one loop results of the CFT to spatial
infinity, consistency conditions arise. This requires that four D-6-branes
sit at each of the two orientifold 6-planes resulting in an 
$SO(8)\times SO(8)$ gauge symmetry in the effective six dimensional
theory. Perhaps we should emphasize that this is the first attempt
at carrying out the orientifold construction explicitly in a curved background. 
It would be desirable to understand some of the subtleties that arise in dealing 
with nontrivial background better. Especially the role of the space dependent 
dilaton field deserves further study.  

\bigskip

\noindent{\bf Acknowledgement:} We are grateful to Stefan Theisen for
extensive discussion. The work of S.F.\ is supported by GIF, the
German Israeli Foundation for Scientific Research. The research of
D.G.\ is supported by the Alexander von Humboldt Foundation. S.P.\
is grateful to the Physics Department of Universit\" at M\" unchen,
and particularly Stefan Theisen for hospitality during the course of
this work. The work presented here is supported in part by TMR 
program ERBFMX-CT96-0045. 


\end{document}

%% file: fig.pstex_t
\begin{picture}(0,0)%
\epsfig{file=fig.pstex}%
\end{picture}%
\setlength{\unitlength}{0.00041700in}%
\begingroup\makeatletter\ifx\SetFigFont\undefined
\def\x#1#2#3#4#5#6#7\relax{\def\x{#1#2#3#4#5#6}}%
\expandafter\x\fmtname xxxxxx\relax \def\y{splain}%
\ifx\x\y   
\gdef\SetFigFont#1#2#3{%
  \ifnum #1<17\tiny\else \ifnum #1<20\small\else
  \ifnum #1<24\normalsize\else \ifnum #1<29\large\else
  \ifnum #1<34\Large\else \ifnum #1<41\LARGE\else
     \huge\fi\fi\fi\fi\fi\fi
  \csname #3\endcsname}%
\else
\gdef\SetFigFont#1#2#3{\begingroup
  \count@#1\relax \ifnum 25<\count@\count@25\fi
  \def\x{\endgroup\@setsize\SetFigFont{#2pt}}%
  \expandafter\x
    \csname \romannumeral\the\count@ pt\expandafter\endcsname
    \csname @\romannumeral\the\count@ pt\endcsname
  \csname #3\endcsname}%
\fi
\fi\endgroup
\begin{picture}(9312,7224)(301,-8773)
\put(3226,-4936){\makebox(0,0)[lb]{\smash{\SetFigFont{6}{7.2}{rm}D6-brane}}}
\put(3226,-5461){\makebox(0,0)[lb]{\smash{\SetFigFont{6}{7.2}{rm}Orientifold-6-plane}}}
\put(301,-5761){\makebox(0,0)[lb]{\smash{\SetFigFont{6}{7.2}{rm}sits here}}}
\put(301,-4711){\makebox(0,0)[lb]{\smash{\SetFigFont{6}{7.2}{rm}NS-5-brane}}}
\put(8776,-2161){\makebox(0,0)[lb]{\smash{\SetFigFont{6}{7.2}{rm}flat space}}}
\put(8776,-1861){\makebox(0,0)[lb]{\smash{\SetFigFont{6}{7.2}{rm}Asymptotically}}}
\put(4201,-2161){\makebox(0,0)[lb]{\smash{\SetFigFont{6}{7.2}{rm}Domain}}}
\put(4201,-1861){\makebox(0,0)[lb]{\smash{\SetFigFont{6}{7.2}{rm}CFT}}}
\put(901,-2161){\makebox(0,0)[lb]{\smash{\SetFigFont{6}{7.2}{rm}Domain}}}
\put(901,-1861){\makebox(0,0)[lb]{\smash{\SetFigFont{6}{7.2}{rm}Strong coupling}}}
\end{picture}